\documentclass[aip,jcp,reprint]{revtex4-1}

\usepackage[pdftex]{hyperref}
\usepackage{graphicx}
\usepackage{amssymb}
\usepackage{amsmath}
\usepackage{verbatim}
\usepackage{bm}
\usepackage{bbold}
\usepackage[separate-uncertainty=true,multi-part-units=single]{siunitx}
\usepackage{color}
\DeclareSIUnit\micron{\micro\metre}

\newcommand{\FFF}{\mathcal{F}}
\newcommand{\XXX}{\mathcal{X}}
\newcommand{\FF}{\boldsymbol{F}}
\newcommand{\TT}{\boldsymbol{T}}
\newcommand{\del}{\partial}
\newcommand{\RRR}{\mathcal{R}}
\newcommand{\VV}{\mathcal{V}}

\begin{document}


\title{Efficient shapes for microswimming: from three-body swimmers to helical flagella}


\author{Bram Bet}
\email[]{B.p.bet@uu.nl}
\affiliation{Institute for Theoretical Physics, Center for Extreme Matter and Emergent Phenomena, Utrecht University, Princetonplein 5, 3584 CC Utrecht, The Netherlands}

\author{Gijs Boosten}
\affiliation{Debye Institute for Nanomaterials Science, Utrecht University, Princetonplein 1, 3584 CC  Utrecht ,The Netherlands}
\author{Marjolein Dijkstra}
 \affiliation{Debye Institute for Nanomaterials Science, Utrecht University, Princetonplein 1, 3584 CC  Utrecht ,The Netherlands}
%
\author{Ren\'e van Roij}
\affiliation{Institute for Theoretical Physics, Center for Extreme Matter and Emergent Phenomena, Utrecht University, Princetonplein 5, 3584 CC Utrecht, The Netherlands}



\date{\today}

\begin{abstract}
We combine a general formulation of microswimmmer equations of motion with a numerical bead-shell model to calculate the hydrodynamic interactions with the fluid, from which the swimming speed, power and efficiency are extracted. From this framework, a generalized Scallop Theorem emerges. The applicability to arbitrary shapes allows for the optimization of the efficiency with respect to the swimmer geometry. We apply this scheme to `three-body swimmers' of various shapes and find that the efficiency is characterized by the single-body friction coefficient in the long-arm regime, while in the short-arm regime the minimal approachable distance becomes the determining factor.
Next, we apply this scheme to a biologically inspired set of swimmers that propel using a rotating helical flagellum. Interestingly, we find two distinct optimal shapes, one of which is fundamentally different from the shapes observed in nature (e.g. bacteria).
\end{abstract}

\pacs{}

\maketitle

\section{Introduction}
For many organisms, motility is of vital importance to survive, since it enables them to search for food or escape from predators. Motility of microorganisms in a fluid takes the form of swimming, where they can often orient themselves toward sources of nutrition, light or the direction of gravity \cite{elgeti2015physics,poon2013clarkia}.
Nature displays a large variety of ways in which microorganisms achieve locomotion. Some organisms propel using rotating helical shaped flagella, such as Escherichia coli \cite{Darnton:2007aa,berganderson,silverman}, use flexible flagella that beat in wave-like patterns such as sperm cells \cite{Alvarez2014198,fisher2014dynamics}, or utilize a large number of cooperatively beating cilia on their surface to propel \cite{elgeti2015physics}.
Also, locomotion of synthetic swimmers or robots is a well-studied subject, with possible applications in efficient drug delivery in the body \cite{peyer2013bio,iacovacci2015untethered}. Theoretically, many designs were proposed. Purcell \cite{purcell} proved that a swimmer with a single internal degree of freedom cannot achieve a net propulsion and proposed the next-simplest design: the `three-link swimmer' \cite{Becker:2003}, which has also been experimentally realized \cite{chan2009bio}. Golestanian et al. \cite{Najafi:2004, Golestanian:2008} proposed another simple swimmer with two degrees of freedom: the `three-bead swimmer', which was studied and generalized extensively \cite{Earl:2007,Alexander:2009,ten2015can,felderhof2014swimming,Vladimirov:2013,Avron:2005}. 
Another strategy towards synthetic microswimmers is to imitate biological swimmers, where examples include swimmers with flagella that perform beating or rotating strokes \cite{dreyfus,Keaveny:2008,tierno2008magnetically,gadelha2013optimal,williams2014self}, or make use of helical structures for propulsion \cite{Zhang,zhang2009characterizing,ghosh2009controlled,keaveny2013optimization,SwanBrady2011,Scherr:2015,Rodenborn:2013}.
\begin{figure}[b!]
   \centering
 \includegraphics[width=0.48\textwidth]{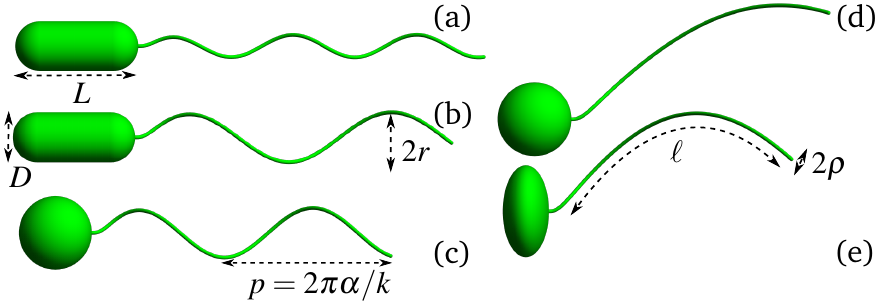}
   \caption{(color online) E. coli-inspired swimmers (a)-(e) with a cell body of length $L$ and diameter $D$  connected to a rigid flagellum of length $\ell$ and radius $\rho$ of helical shape with radius $r$ and pitch $p$.}
   \label{fig:helixswimmers}1
\end{figure}

In many designs of artificial swimmers, the propulsion is driven by an internally stored, and therefore limited, fuel supply. Therefore, when optimizing the swimmer design, one should take into account the swimming efficiency rather than the swimming velocity. The efficiency is defined as the ratio between the propulsion power and the total dissipated power, such that higher efficiencies are associated to swimmers that move faster at the same fuel consumption. 
In this article, we combine a general formulation of low-Reynolds number swimmer equations of motion with a numerical method based on a bead-shell model \cite{HYDRO,torre} to numerically determine the grand resistance tensor of many-component swimmers, from which we extract the swimming velocity, power and efficiency. The advantage of this method is that it is applicable to any collection of connected rigid objects, while it is also computationally relatively cheap and allows for calculation of the shape-dependent power and efficiency.

\section{Model \& method}
\subsection{Equations of motion}
We consider a swimmer consisting of $N$ parts of a fixed shape, immersed in a quiescent incompressible Newtonian bulk fluid of viscosity $\eta$, without any external body force. For instance, swimmers composed of a rigid head and a rigid tail, such as shown in Fig. \ref{fig:helixswimmers}, are described by $N=2$. We let $\XXX$ denote the $6N$-vector with components $\XXX_i = (\bm{R}_i , \bm{\Theta}_i)$ denoting the positions $\bm{R}_i$ and orientation angles $\bm{\Theta}_i$ with respect to a fixed reference frame of component $i = 1,... , N$, and $\dot{\XXX}$ the corresponding (angular) velocities. The swimmer components are connected by mechanical actuators or motors that impose their relative coordinates $x_i \equiv \XXX_i -\XXX_N$ and velocities $\dot{x}_i$, where we choose the $N$-th part as a reference. This motion induces a fluid flow $\bm{u}$ and pressure field $p$ that give rise to a viscous (friction) force field on the surface of the swimmer, that in turn gives rise to a net displacement of the swimmer. In the low Reynolds number regime, the hydrodynamics is described by the Stokes equation \cite{elgeti2015physics}
\begin{equation} \label{eq:stokes}
 - \nabla p + \eta \nabla^2 \bm{u} =0,  \qquad \nabla \cdot \bm{u} =0 \, ,  
\end{equation}
supplemented with no-slip boundary conditions on the surface of each swimmer part, and vanishing $\bm{u}$ at infinity. Due to the linearity of the Stokes equation, one derives that $\FFF_i = (\FF_i,\TT_i)$, with the forces $\FF_i$ and torques $\TT_i$ acting on the $i$-th swimmer part, relates linearly to the particle (angular) velocities \cite{brenner1964stokes,brenner1972stokes},
\begin{equation} \label{eq:generaleom}
\FFF_i(\XXX,\dot{\XXX}) = - \eta \ \RRR_{ij}(x) \ \dot{\XXX}_j \,,  
\end{equation}
where repeated indices imply summation over the swimmer parts and the $6N \times 6N$ tensor $\RRR$ denotes the grand resistance tensor, which we will calculate below. Due to translational and rotational invariance, this tensor depends only on the relative coordinates $x$, and furthermore on the shape of the different swimmer parts. In absence of external forces (such as gravity or externally applied magnetic fields), the total force  $\sum_{i=1}^N \FF_i $ and torque  $\sum_{i=1}^N \TT_i + \bm{r}_i \times \FF_{i} $ on this swimmer must vanish. Once $\RRR(x)$ is known, the $6$ constraints of the force-free condition, together with the $6N -  6$ constraints $\dot{x}_i$ imposed by the motors, provide enough constraints to solve Eq. (\ref{eq:generaleom}) for $\dot{\XXX}_j$, which for the cases of interest below 
gives\footnote{We find $\bm{r}_i \parallel \FF_i$ for the three-body swimmers and $\bm{r}_i = 0$ for the helical flagellum swimmers, such that $\bm{r}_i \times \bm{F}_i = 0$ and hence $\sum_{i=1}^N \FFF_i=0$, from which Eq. (\ref{eq:velocityN}) follows. In general, the linear relation $\dot{\XXX}_N = \mathcal{V}_j \cdot \dot{x}_j $ stays valid, but the expression for $\mathcal{V}$ is more involved.
}
\begin{equation} \label{eq:velocityN}
\dot{\XXX}_N = -\left(\sum_{k,l=1}^N \RRR_{kl}\right)^{-1} \left( \sum_{i=1}^N \RRR_{ij} \dot{x}_j \right) \equiv \mathcal{V}_j(x) \cdot \dot{x}_j \,,
\end{equation}
from which $\FFF$ follows from inserting this expression into Eq. (\ref{eq:generaleom}). The $6N-$vector field $\mathcal{V}$ expresses the linear coupling of the motor-imposed velocities $\dot{x}_j$ to the motion of our arbitrarily chosen reference part.
To calculate the displacement $\bm{\Delta}$ per stroke, it is sufficient to consider the displacement $\int_0^T dt \, \dot{\XXX}_N$ of component $N$, since the internal coordinates $x$ vary cyclically during a stroke of period $T$. Hence,
\begin{equation} \label{eq:displacement}
\bm{\Delta} =\! \int_0^T \! dt \, \dot{x}_j \cdot \mathcal{V}_j(x(t)) =  \oint_{\del \Sigma} dx_j \cdot \mathcal{V}_j =\! \int_{\Sigma} d (dx_j \cdot \mathcal{V}_j),\! \!
\end{equation}
where $\del \Sigma$ is a closed path enclosing an area $\Sigma$ in the $(6N-6)$-dimensional internal coordinate space that describes the swimming 
stroke. Note that $\Sigma$ can not be defined if there is only a single degree of freedom that is rotational and describes a $2\pi$ rotation; in this case the displacement should be calculated by the contour integral.
In the second equality we used that $\dot{x}_j dt = dx_j$ and in the last equality we used the (generalized) Stokes Theorem, where the operator $d$ on the right hand side denotes the so-called exterior derivative: $d (dx_j \cdot \mathcal{V}_j) = \del_k \VV_j \ d x_k \wedge d x_j$ \cite{frankel2011geometry}.  Eq. (\ref{eq:displacement}) is a general formulation of the Scallop theorem\cite{purcell}: a reciprocal stroke is one that does not enclose any area, such that the displacement vanishes. We define the average swimming velocity 
\begin{equation}
\langle \bm{U} \rangle = \bm{\Delta} / T,
\end{equation} 
and a generalized swimming or Lighthill efficiency \cite{Shapere:1989aa,Felderhof:1994aa,Golestanian:2008} as 
\begin{equation} \label{eq:efficiency}
\eta_L = \frac{\langle \dot{\XXX}_i \rangle  \langle \RRR_{ij} \rangle \langle \dot{\XXX}_j \rangle}{ \langle \dot{\XXX}_i \RRR_{ij}  \dot{\XXX}_j \rangle} =  \frac{ \langle \bm{U} \rangle \cdot \eta \langle \bm{\mathfrak{R}} \rangle \cdot \langle \bm{U} \rangle}{\langle P \rangle},
\end{equation}
where $\langle \cdot \rangle$ denotes the time average over one period, and
\begin{align}
\langle P \rangle &= \frac{-1}{T} \int_0^T dt \ \FFF_i \cdot \dot{\XXX}_i,  \\
\langle \bm{\mathfrak{R}} \rangle  &\equiv \frac{1}{T} \int_0^T dt \ \sum_{i,j=1}^N  \RRR_{ij}, 
\end{align}
denote the time-averaged dissipated power and the effective $6 \times 6$ rigid body resistance tensor, respectively.

\subsection{Numerical methods}
For a swimmer of a certain geometry, we determine the grand resistance tensor $\RRR(x)$ using a bead-shell model \cite{torre}. In the conventional implementation of this model, the surface of a rigid particle ($N=1$) is covered by $M \gg 1$ spheres, whose radius $a$ is small compared to the size $R$ of the particle. 
These spheres are distributed (quasi-)homogeneously on the surface, which we achieve here using a simulated-annealing method. In this method, the spheres are given a repulsive interaction and are stochastically moved on the surface according to the Metropolis algorithm. The temperature that appears in the Boltzmann factors dictates the acceptance and rejection and is slowly lowered to find a near-homogeneous distribution of the spheres on the surface.

When given a finite common velocity $\bm{V}$, the induced flow field causes pair interactions between the little spheres, given by the Rotne-Prager mobility tensor\cite{rotneprager,Wajnryb2013} $\bm{\mu}^{\text{RP}}_{kl}$ as $\bm{V}= \bm{V}_k = \sum_{l=1}^M \bm{\mu}^{\text{RP}}_{kl} \bm{F}_l$, with $\bm{V}_k$ and $\bm{F}_k$ the velocity of, and force on sphere $k$, respectively, and  
\begin{equation}
\bm{\mu}_{kl}^{RP} = \frac{1}{8\pi \eta r_{kl}} \left( \left[1 + \frac{2 a^2}{r_{kl}^2} \right] \mathbb{1}_3 +  \left[1 - \frac{2 a^2}{r_{kl}^2} \right] \frac{\bm{r}_{kl}  \bm{r}_{kl}}{r_{kl}^2}  \right),
\end{equation}
with $\bm{r}_{kl} = \bm{r}_k - \bm{r}_l$. 
The forces on each sphere can then be calculated by $3M\times3M$ matrix inversion, $\bm{F}_k = \sum_{l=1}^M ((\bm{\mu}^{\text{RP}})^{-1})_{kl} \bm{V}_{l}$, from which the total force $\bm{F}$ and torque $\bm{T}$ on the rigid object follow as the sum of the individual forces and torques around a chosen reference point $\bm{r}_O$ (i.e. center of mass): $\bm{F} = \sum_{k=1}^M \bm{F}_k$ and $ \bm{T} = \sum_{k=1}^M \bm{F}_k \times (\bm{r}_k- \bm{r}_O)$. In this work, the matrix inversion is done using an LU factorization routine of the LAPACK package \footnote{See http://www.netlib.org/lapack/ for information and downloadable versions of LAPACK}. Subsequently, we determine the resistance tensor $\RRR$ for an increasing number (typically $1000-3000$) of spheres of decreasing size, while keeping the total bead surface $4\pi a^2 M$ constant and equal to the surface area of the body of interest. Next, a quadratic function of $a$ is fitted to the results for each component of $\RRR$, after which the limit $M \to \infty$ is determined by the intersect at $a/R=0$. By taking the limit with this specific surface coverage, we retrieve the boundary integral formulation of the Stokes equation \cite{kim1991,guazzelli2011physical}, guaranteeing accurate results for $\RRR$. Note that the contribution of the torques on the individual beads to the total torque can be neglected, as it vanishes in the limit $a \to 0$. Therefore we only consider the forces $\bm{F}_k$ and not the torques $\bm{T}_k$ on the beads, from which we achieve a decrease in computation time ($3M\times3M$ versus $6M\times6M$ matrix inversion). 

\begin{figure}[t]
\includegraphics[width=0.5\textwidth]{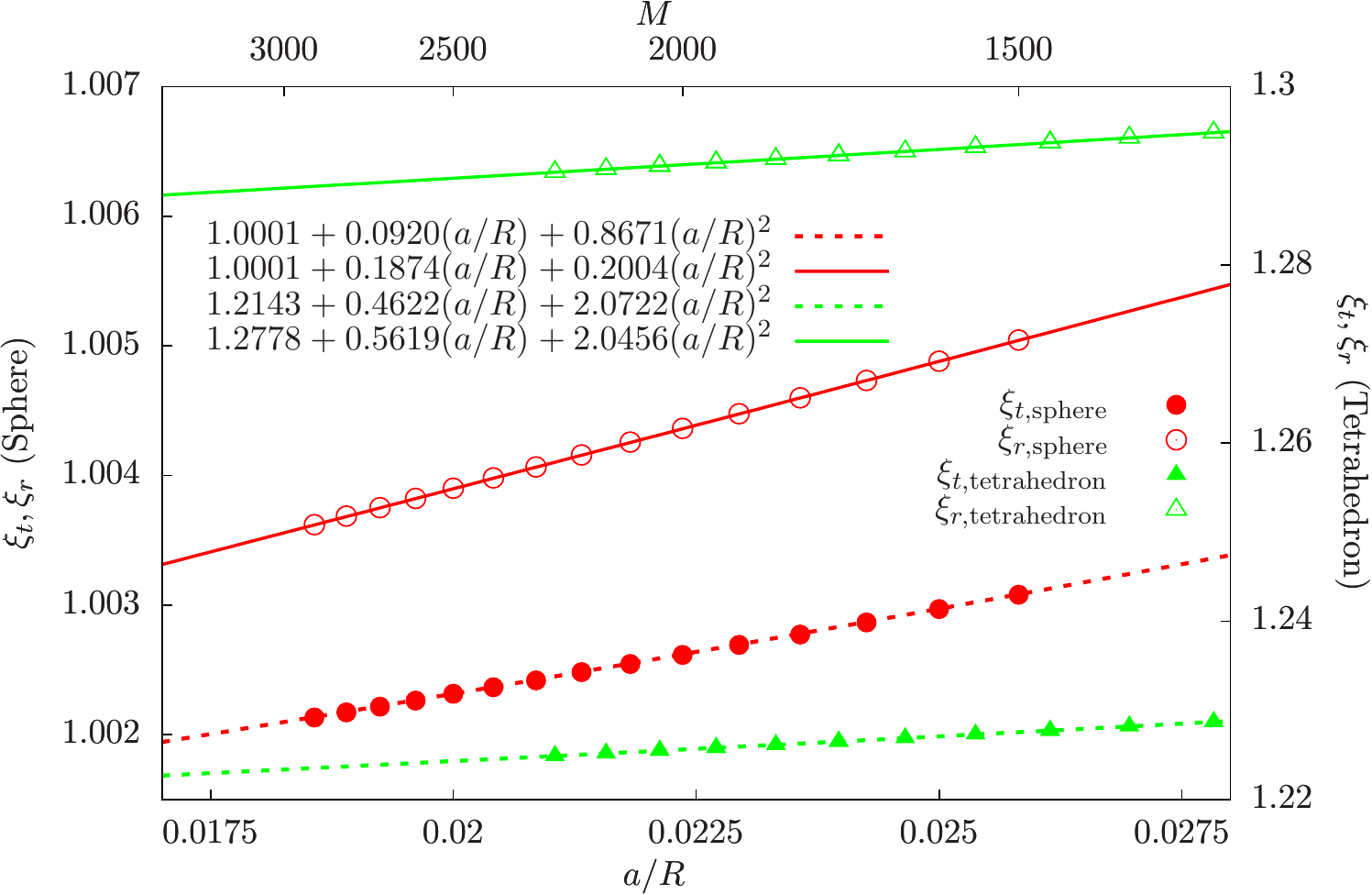}
 \caption{Friction coefficients $\xi_t,\xi_t$ as a function of the relative bead size $a/R$, for translational (red dots) and rotation(red circles) of a sphere, and the corresponding quadratic fits (dashed and full red line, respectively). The values of the fit coefficients are shown in the legend, the final calculated friction coefficient is determined by setting $a/R=0$. In green we show similar results for the tetrahedron. } \label{fig:convergence}
\end{figure}

In Fig. \ref{fig:convergence} we show a few illustrative results of this extrapolation. Firstly, we show results for the translational friction coefficient $\xi_t$ (red dots) and rotational friction coefficient $\xi_r$ (red circles) for a sphere of radius $R$, as obtained from the bead-shell computation as function of relative bead size $a/R$. These quantities express the translational and rotational friction of a rigid body, in units of $6\pi \eta R$ and $8 \pi \eta R^3$, respectively, which are the analytically known results for a sphere of radius $R$. Precise definitions of $\xi_t$ and $\xi_r$ are given in the next section. The quadratic fits are indicated by the dashed (translation) and full (rotation) red line in Fig. \ref{fig:convergence}, while the best-fit values of the coefficients are also indicated. The small-$a$ limit can be compared to the exact theoretical value of $1$ and therefore serves as an estimate of the accuracy of this model. We observe that the error is of the order of $10^{-4}$. The green curves in Fig. \ref{fig:convergence} show $\xi_t$ and $\xi_r$ for the tetrahedron of volume $\frac{4 \pi}{3} R^3$, with numerical values on the right vertical axis, where the full triangles and dashed line correspond to translation and the open triangles and full line correspond to rotation. Again, the values of the fit coefficients can be seen in the legend, the small-$a$ limit being the result of interest. 

We extend this bead-shell model to allow for non-rigid objects with internal degrees of freedom ($N>1$). The swimmer surface is again covered with a large number of spheres $M$, distributed over the $N$ different components. Next, we impose a non-zero relative velocity between the components and solve for the hydrodynamic force on each of the components, constructing the full tensor $\RRR$ in this way. In principle, one needs to do this calculation for a (large) number of internal configurations $x$ along the path $\del \Sigma$ in the $6N-6$ dimensional configuration space in order to evaluate Eq. (\ref{eq:displacement}) numerically.

The main advantage of this bead-shell method is that it allows for accurate results with relative short computation time: the calculation of the $6\times6$ resistance tensor of a sphere with a relative precision of the order of $10^{-4}$ with respect to the exact results, takes only a few minutes on a desktop computer. The calculation time for any other rigid body is similar. In contrast, a full three-dimensional finite element calculation would take considerably longer when the shape under consideration does not allow for simplifications due to symmetry. Specifically for determining the (swimmer) resistance tensor, the bead-shell method benefits from the fact that for a single surface-covering configuration of spheres, after the many-sphere Rotne-Prager mobility tensor is LU factorized, each component of the resistance tensor is calculated very quickly by iterating over the different degrees of freedom (rigid body or internal). For a finite-element method on the other hand, this would amount to calculating the full velocity profile for a different set of boundary conditions which is therefore computational much more costly.

\section{Results}
\subsection{Rigid bodies: platonic solids}
For $N=1$, $\RRR$ is the resistance tensor of a single rigid body. As a proof of concept, we use the bead-shell model to calculate $\RRR$ for each of the five platonic solids, which possess sufficient symmetry for the resistance tensor to be isotropic, characterized by the two dimensionless friction coefficients $\xi_t$ and $\xi_r$ for translation and rotation, defined by 
\begin{equation}
\eta \RRR = (6 \pi \eta R \xi_t) \mathbb{1}_3 \oplus (8 \pi \eta R^3 \xi_r^3) \mathbb{1}_3.
\end{equation} 
Here, $\mathbb{1}_3$ is the three-dimensional unit matrix and $R = ( 3V/4\pi)^{1/3}$ is the effective radius in terms of the particle volume $V$, such that $\xi_t = \xi_r = 1$ for a sphere. 
\begin{table}[t] \label{tab:frictioncoeff}
\centering
\begin{tabular}{l|llllll}
        & \ \includegraphics[width=0.6cm]{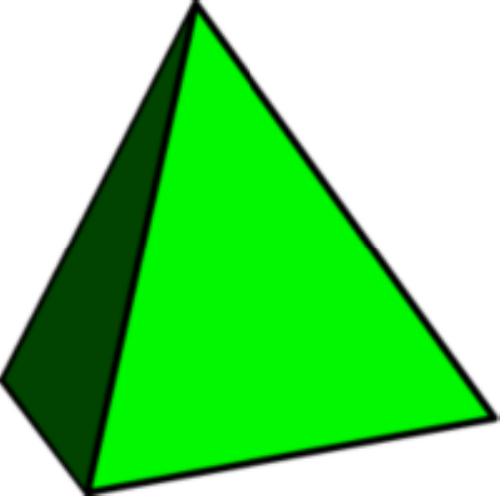} \quad 
     & \ \includegraphics[width=0.6cm]{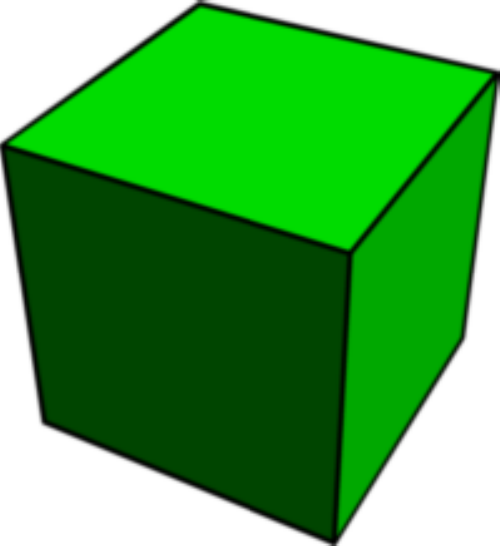}   \quad 
        &\ \includegraphics[width=0.6cm]{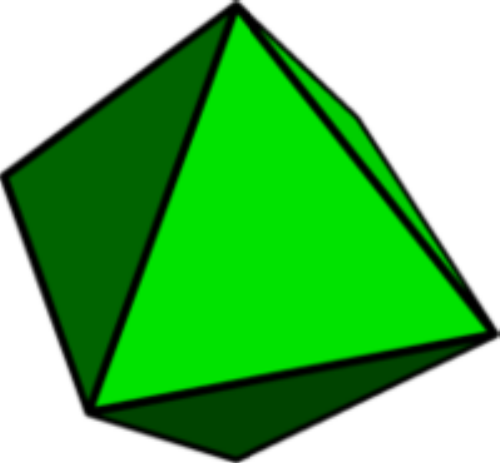}     \quad 
          &\ \includegraphics[width=0.6cm]{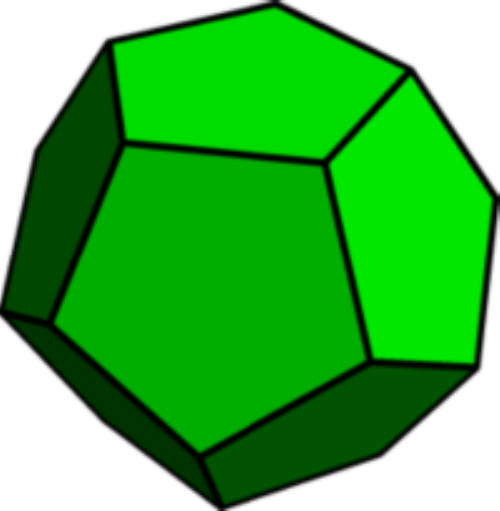}     \quad 
            &\ \includegraphics[width=0.6cm]{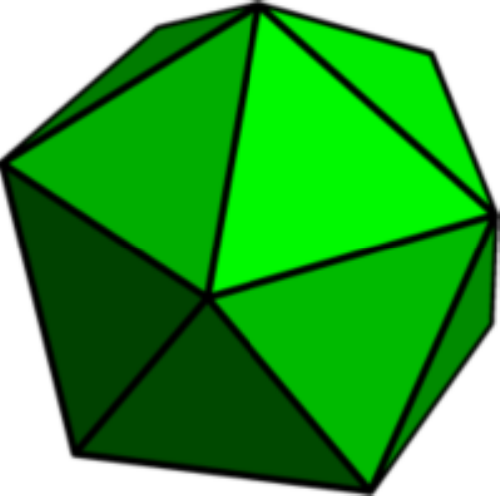}    \quad 
                &\ \includegraphics[width=0.6cm]{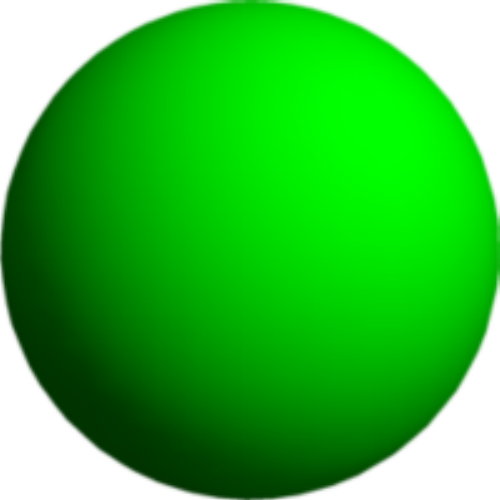}       \\ \hline
$n_f$   &  $4$       & $6$       &$ 8 $      & $12$      & $20$      & $-$ \\
$\xi_t$ & $1.214$ & $1.086$ & $1.072$ & $1.027$ & $1.019$ & $1.000$      \\
$\xi_r$ & $1.278$ & $1.102$ & $1.089$ & $1.030$ & $1.022$ & $1.000$ \\
$\sqrt{A/4\pi R^2}$ & $1.221$ & $1.114$ & $ 1.087$ & $1.048$ & $1.032$ & $1.000$
\end{tabular}
\caption{Number of faces $n_f$ and relative translation and rotation friction coefficients, $\xi_t$ and $\xi_r$ respectively, of the five platonic solids (tetrahedron, cube, octahedron, dodecahedron and icosahedron) and the sphere. In the bottom row the ratio between the surface-area-defined length scale $\sqrt{A/4\pi}$ and the volume-defined unit of length $R=(3V/4\pi)^{1/3}$ is shown.} \label{tab:frictioncoeff}
\end{table} 
In Table \ref{tab:frictioncoeff} we list $\xi_t$ and $\xi_r$ and observe that $\xi_r > \xi_t > 1$ for all five platonic solids, the more so for bodies with fewer faces, with enhanced friction compared to the sphere of equal volume exceeding $20\%$ for the tetrahedron. Since $\xi_r > \xi_t$, it is impossible to assign a single hydrodynamic radius to any of the platonic solids, the translational radius $\xi_t R$ is always smaller than the rotational radius $\xi_r R$. 

Interestingly, it turns out that the friction coefficients can qualitatively, and to some extent quantitatively, be estimated by another length scale that is defined by $\sqrt{A/4\pi}$, where $A$ is the surface area of the solid body of interest. Specifically, we consider this length scale in units of  the volume-defined unit length $R = ( 3V/4\pi)^{1/3}$ in the bottom row of Table \ref{tab:frictioncoeff} and observe that it agrees approximately with the calculated friction coefficients:
\begin{equation} \label{eq:frictioncoeffestimate}
\xi_i \approx  \sqrt{\frac{A}{4\pi R^2}} = \frac{ A^{1/2}}{6^{1/3} \pi^{1/6} V^{1/3} } \approx 0.455 \ A^{1/2} / V^{1/3},
\end{equation}
where $\xi_i$ denotes either $\xi_t$ or $\xi_r$. Alternatively, one can formulate this estimate in terms of the hydrodynamic radius as $R_h = \xi_t R \approx \sqrt{A/4\pi}$. Obviously, this relation is not exact and does not distinguish between translational and rotational friction, but it may serve as an estimate for experimental purposes where both the volume and surface area of a particle are known. One should also note that this estimate breaks down for particles with resistance tensors that are strongly anisotropic. For example prolate ellipsoids of large aspect ratio, where the rotational friction factors in different directions differ over orders of magnitude and can therefore not be accurately estimated by Eq. \ref{eq:frictioncoeffestimate}, which is easily checked with the known exact friction coefficients\cite{Perrin2}.   

\subsection{Three-body swimmers}
\begin{figure}[b!]
\includegraphics[width=0.45\textwidth]{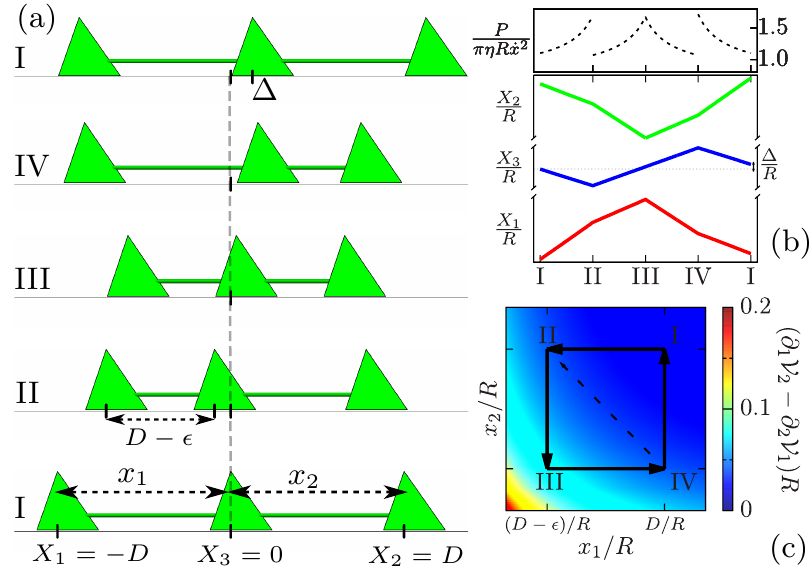}
 \caption{(color online) Fourfold stroke of a three tetrahedron-swimmer (a). Positions $X_i(t)$ of the individual parts and instantaneous power $P(t)$ during a stroke (b). Stroke path $\del \Sigma$ represented in $x$-space, with a density plot of $d(\VV_i dx_i)$ (c).}  \label{fig:3b_stroke}
\end{figure}

\begin{figure}[tbp]
   \centering
   \includegraphics[width=0.5\textwidth]{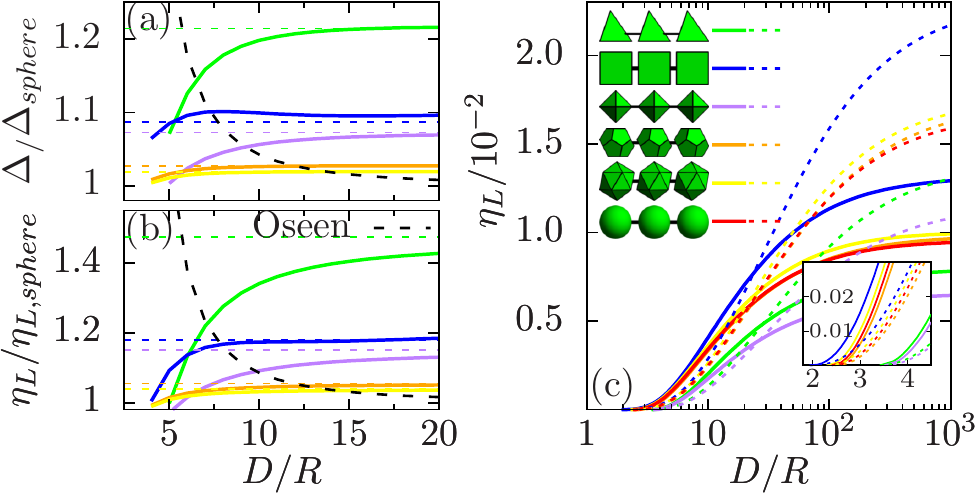}
   \caption{(color online) Ratio of displacement (a) and efficiency (b) of the three-platonic solid swimmer compared to the three-sphere swimmer as a function of maximum arm length $D$, for amplitude $\epsilon/R = 1$. The horizontal dashed lines indicate $\xi_t$ (a) and $\xi_t^2$ (b). Efficiency $\eta_L$ of a stroke with $D-\epsilon$ close to contact, as a function of $D$, for a square stroke I-II-III-IV-I (full lines) and a triangular stroke II-III-IV-II (dashed lines) (c). The inset shows a zoom for small $D/R$. The legend of (c) illustrates the relative orientations of the three bodies and the fixed minimal separation.}
   \label{fig:3b_results}
\end{figure}

One of the simplest swimmers that can be described by our new method is composed of $N=3$ rigid bodies connected by two arms of time-dependent lengths $x_i(t)$ driven by a motor. Earlier works on this three-body swimmer mainly consider a three-sphere set-up, with hydrodynamics modeled by the Oseen tensor that is only accurate in the regime of long arms and small spheres. In this work, by making use of a bead-shell model to determine the resistance tensor, we do not suffer from these restrictions.  In Fig. \ref{fig:3b_stroke}(a), the swimmer design and stroke cycle I-II-III-IV-I are illustrated for a swimmer consisting of three tetrahedra. The stroke is performed by periodically and non-reciprocally changing $x_i(t)$ between a maximum $D$ and a minimum $D-\epsilon$, causing the swimmer to go back and forth, resulting in a displacement $\Delta$ after one period. The positions $X_i(t)$ of the individual parts and the instantaneous power $P(t)$ during the stroke are illustrated in Fig. \ref{fig:3b_stroke}(b). In Fig. \ref{fig:3b_stroke}(c) we show a stroke represented as a closed path $\del \Sigma$ in the two-dimensional internal coordinate space $(x_1,x_2)$, where the density plot represents $d(\VV_j dx_j) = (\del_1 \VV_2 - \del_2 \VV_1) dx_1 dx_2$ (see Eq. (\ref{eq:displacement})). This function is strictly positive and decreases with $x_1$ and $x_2$, implying that the displacement per stroke decreases with $D$ (for fixed $\epsilon$) and increases with $\epsilon$. As the platonic solids do not posses the full spherical geometry, there are obviously many possible (relative) orientations of the three bodies. To avoid ambiguity, we only show results for three-body swimmers with one and the same fixed orientation of all three components with respect to the axes that connects the three bodies, as indicated by the legend in Fig. \ref{fig:3b_results}. We point out that the results do not differ significantly for other cases.
Animations of three-body swimmers can be found in the supplementary material.

In Fig. \ref{fig:3b_results} we show the displacement (a) and efficiency (b) of three-body swimmers consisting of each of the platonic solids as a function of the maximum arm length $D$ for fixed small amplitude $\epsilon = R$, compared to a three-sphere swimmer performing the identical stroke. We observe from Fig. \ref{fig:3b_results}(a) that in the regime of long arms, the displacement ratio $\Delta/\Delta_{\text{sphere}}$ tends to the friction coefficient $\xi_t$, indicated by the horizontal dashed lines. This result is consistent with that of Earl et al. \cite{Earl:2007}, who showed for three-sphere swimmers that $\Delta \propto R$ for the Oseen interaction in the long arm regime, if we take $\xi_t R \equiv R_h$ as the hydrodynamic radius. In Ref. \cite{Earl:2007} it is also shown that the instantaneous velocity of sphere-swimmers is independent of $R$ for large $D$, which for general swimmers also holds true as can be deduced from Eq. (\ref{eq:velocityN}). 
Furthermore, to leading order in $R/D$ the individual forces $F_i$ and the average rigid body friction tensor $\langle \bm{\mathfrak{R}} \rangle$ are both proportional to $R_h$, such that $P \propto R_h$ and hence we find that $\eta_L \propto \xi_t^2$ for large $D$, which is precisely what is observed in Fig. \ref{fig:3b_results}(b). Concluding, we observe that particles with larger friction constitute more efficient swimmers, which is interesting given the fact that the opposite holds for externally driven (e.g. sedimenting) particles, where particles that experience more friction move slower. The black dashed lines in Fig. \ref{fig:3b_results}(a) and \ref{fig:3b_results}(b) show a comparison with the Oseen interaction model used by Najafi and Golestanian \cite{Najafi:2004, Golestanian:2008}, where we observe an agreement in the long-arm regime, but a qualitatively erroneous trend in the short-arm regime. This discrepancy is explained by the breakdown of the Oseen approximation at small distances, while the bead-shell model extrapolates to infinitesimal bead size, such that it holds up to distances comparable to the used bead sizes, which are two orders of magnitude smaller than the rigid bodies under consideration.

We explore the maximum efficiency by considering strokes from $D-\epsilon$ close to contact to a maximum separation $D$ in Fig. \ref{fig:3b_results}(c). Here, we define the minimum gap as being $20\%$ of the center-to-center distance at which the bodies start to overlap. This nearest distance depends on the shape and orientation, as for instance two cubes oriented with the faces adjacent can approach each other more closely than two tetrahedra in this particular orientation. This minimal separation is illustrated in the legend of Fig. \ref{fig:3b_results}(c).  Indeed, for this maximum-amplitude stroke we observe that the most efficient swimmer is the three-cube swimmer, which can attain the smallest contact distance.

Note that in this analysis, we focussed on the effect of the body shape on the efficiency for a given prescribed stroke, rather than optimizing the stroke itself as is done for instance in Ref. \cite{alouges2008optimal}, where the instantaneous power is kept constant during the stroke. We find that our results for a reparametrization of the stroke that fixes the power rather than the internal velocity differ negligibly from the results presented in Fig. \ref{fig:3b_results}. Since the resistance tensor depends on the internal configuration of the swimmer, even when keeping the internal velocities constant in time, the forces and therefore the instantaneous power will vary with time. On the other hand, demanding that the power is constant in time will require adjusting the internal velocities in a nonlinear fashion in time. We also point out that a representation of the displacement and efficiency in terms of the amplitude $\epsilon$ rather than the maximum arm length $D$ (in both cases keeping the minimum arm length $D-\epsilon$ fixed to near contact) gives rise to qualitatively identical results.

From the color map of Fig. \ref{fig:3b_stroke}(c), we observe that the upper triangle I-II-IV-I contributes less to the displacement than the lower triangle II-III-IV-II, while we observe from Fig. \ref{fig:3b_stroke}(b) that the (average) power in all four steps is comparable. Therefore, we propose the new stroke II-III-IV-II, where both motors operate at the same time while crossing the diagonal IV-II. The $D$-dependence of the efficiency of such a stroke, with $D-\epsilon$ close to contact, is shown by the dashed lines in Fig. \ref{fig:3b_results}(c). We see that the square stroke is yet favorable for small amplitudes, while for larger $D$ the triangular stroke becomes much more efficient by a factor $\sim 2$, with a crossover regime around $D/R \approx 20$. Also for this triangular stroke, the differences in efficiency for the various shapes are determined by the smallest contact distance, as can be seen in the inset of Fig. \ref{fig:3b_stroke}(c).

\subsection{Helical flagellum swimmers}
Our numerical method is not limited to relatively simple designs such as the three-body swimmer. We turn the discussion to a swimmer that propels itself by a rotating helical flagellum, not unlike E. coli bacteria \cite{Darnton:2007aa,berganderson,silverman}. 

We assume this swimmer to consist of two parts, an axially symmetric cell body and a helical flagellum, that can rotate with respect to each other. The helical flagellum is attached to the surface of the cell body in such a way that the center of this attachment lies in the origin of the defined coordinate system $\{ \hat{\bm{x}}, \hat{\bm{y}}, \hat{\bm{z}} \}$. The centerline of the helical shape of contour length $\ell$ is parametrized for $s \in (0,\ell)$ by 
\begin{equation} \label{eq:centerline}
\bm{h}(s) = r f(s) \left(\cos(k s) \, \hat{\bm{x}} + \sin(ks) \, \hat{\bm{y}} \right) + \alpha s \, \hat{\bm{z}},
\end{equation}
with $ \alpha^2 + r^2 k^2 = 1$, for pitch parameter $\alpha$, radius $r$ and wave number $k$. Here, the function  $ f(s) = s^2 /(s^2 + (c \ell)^2)$ ensures the perpendicular attachment to the surface of the cell body for $c>0$ and asymptotes rapidly to unity for small enough $c$. We find our results to be fairly independent of $c$ in a range of $0.02 < c < 0.1$ and therefore we have fixed $c = 0.05$. In this parameterization, the helical pitch is expressed as $p = 2\pi \alpha / k$. Given the centerline parametrization (\ref{eq:centerline}), the surface of the helical flagellum is parametrized as 
\begin{equation}
\bm{H}(s,\phi) = \bm{h}(s) + \rho \, ( \cos(\phi) \, \bm{n}(s) + \sin(\phi) \,  \bm{m}(s) ),
\end{equation}
where $\boldsymbol{n}(s), \boldsymbol{m}(s)$ are mutually orthogonal unit vectors that are also orthogonal to $d \boldsymbol{h}(s)/ds$. The volume $V$ of the axially symmetric cell body is kept constant for every aspect ratio $L/D$, a fixed unit length is defined by $R = (3 V/ 4 \pi)^{1/3}$ as before.
The swimmer and its relevant shape parameters are shown in Fig. \ref{fig:helixswimmers}, animations of the motion of this type of swimmer can be found in the supplementary material. The relative rotation rate $\dot{\theta}$ between the cell body and the flagellum is imposed by a motor, whereas the other $6$ degrees of freedom of this swimmer are those of a rigid body.  

Analytical\footnote{There are also hydrodynamic simulation studies that capture the dynamics of the E. coli bacterium accurately. \cite{hu2015modelling,watari2010hydrodynamics,kong2015bead}} studies of such helical flagella swimmers usually ignore the rotational asymmetry of the helix around the long axis, and hence its transverse translation and rotation, such that the number of degrees of freedom of the swimmer reduces to three. Moreover, the hydrodynamic interactions between the cell body and the flagellum (the off-diagonal blocks $\RRR_{i \neq j}$) are usually ignored \cite{Lauga2009,poon2013clarkia,elgeti2015physics,Chattopadhyay12092006}. In this study, we do not ignore these features, which turn out to play an important role in certain shape regimes. We do assume the flagellum to be rigid and to retain its shape during the swimming motion, a safe assumption for artificial swimmers which also seems to hold for several biological flagellum swimmers such as E. coli \cite{berganderson,silverman,Darnton:2007aa}. Note that this implies that we neglect any effects of elasticity of the helical filament.

In order to compare velocities and rotation rates of real E. Coli with those predicted by our model, we insert typical shape parameters as reported in \cite{Darnton:2007aa}, $D = \SI{0.88}{ \micron}$, $L = \SI{2.25}{\micron}$, $r = \SI{0.20}{\micron}$, $p= \SI{2.2}{\micron}$, $\ell = \SI{7.1}{ \micron}$ and $\rho = \SI{0.035}{ \micron}$, corresponding to the swimmer shown in Fig. \ref{fig:helixswimmers}(a). Note that E. coli typically have around 10 flagella \cite{Darnton:2007aa,poon2013clarkia} that bundle and synchronize during swimming, which we effectively describe as a single flagellum of approximately three times the filament radius, which is $\SI{0.012}{ \micron} \approx \rho/3$. Also, the reported motor rotation rate equals $\dot{\theta} / 2 \pi = \SI{154}{\hertz}$. 

We find a swimming speed of $v =\SI{17}{ \micron \per \second}$ and body and flagellum rotation rates of $\SI{23}{\hertz}$ and $\SI{131}{\hertz}$, respectively, which should be compared to the observed values $v = \SI{29  +- 6}{\micron \per \second}$, $ \SI{23 +-  8}{\hertz}$ and $ \SI{131+- 31}{\hertz}$ \cite{Darnton:2007aa}. Hence, our method produces fairly accurate results for the complex swimming motion of E. coli. Note that, since the flagellum is not completely rotationally symmetric, the swimming gait shows a periodic transversal `wobble' motion, as can be seen in the animations. However, this `wobble' is smaller than that reported by Ref. \cite{Darnton:2007aa}, which might be explained by the fact that we consider a single flagellum at the polar end of the cell body, rather than several ones attached at several positions.

The numerical values of (some of the) resistance tensor components of this swimmer can be compared to the measurements of Chattopadhyay et al.\cite{Chattopadhyay12092006}, where the components of a three-dimensional resistance tensor were measured for a population of E. coli. We calculated the coefficients for translation of the flagellum along, and rotation around the cell body symmetry axis to be $\SI{0.78 e-8}{\newton \second \per \meter}$ and $\SI{0.99e-21}{\newton \second \meter}$, respectively, while for the cell body these are $\SI{1.0e-8}{\newton \second \per \meter}$ and $\SI{5.5e-21}{\newton \second \meter}$. The off-diagonal component that describes the rotation-translation coupling of the flagellum (again, around the symmetry axis) is $\SI{3.6e-16}{\newton \second}$. We find these values to agree qualitatively with the results in Ref. \cite{Chattopadhyay12092006}, but quantitatively different by $30\%$ to a factor of $2$. These discrepancies can be related to the fact that the measurements in Ref. \cite{Chattopadhyay12092006} are done for a population of E. coli with a spread in shape parameters, as for instance the reported cell body length $L$ varies between 2 and 5 $\si{\micro \meter}$. Indeed, the fact that the measured cell body translational resistance coefficient is higher while the rotational coefficient is lower, indicates that the average cell body aspect ratio of this population was higher than in our calculation\cite{Perrin2}. 

The calculated Lighthill efficiency of this swimmer is $\eta_L = 0.0064 $. A calculation of the efficiency is also done in Ref. \cite{Chattopadhyay12092006}, although a different definition is used where only the cell body translation resistance appears in the numerator of Eq. (\ref{eq:efficiency}). When we correct for this, we find our efficiency to be fourfold lower, which can be traced back directly to the difference in the resistance tensor. Lastly, the power consumed by our swimmer is $\SI{7.8e-16}{\watt}$, which also agrees qualitatively with Ref. \cite{Chattopadhyay12092006} (where it is $\SI{4.3e-16}{\watt}$), but one should keep in mind that this quantity also depends on the motor frequency.

An interesting question, which is of direct relevance for constructing artificial swimmers, is how the efficiency depends on the geometry. One could argue that evolution has selected the most efficient shapes, but also that the efficiency is good enough for the survival of E. coli and that other shapes could be (much) more efficient. Of course, as bacteria use only a fraction of their available energy for swimming \cite{bray2001cell}, other factors than the swimming efficiency could determine the evolutionary fitness. Also, as E. coli perform a run-and-tumble motion\cite{Darnton:2007aa}, the ability to tumble efficiently could also be important. Yet, from the perspective of constructing swimmers with limited internal fuel supply, the geometry-dependent efficiency is an important design feature.
 \begin{figure*}[t]
   \centering
   \includegraphics[width=14cm]{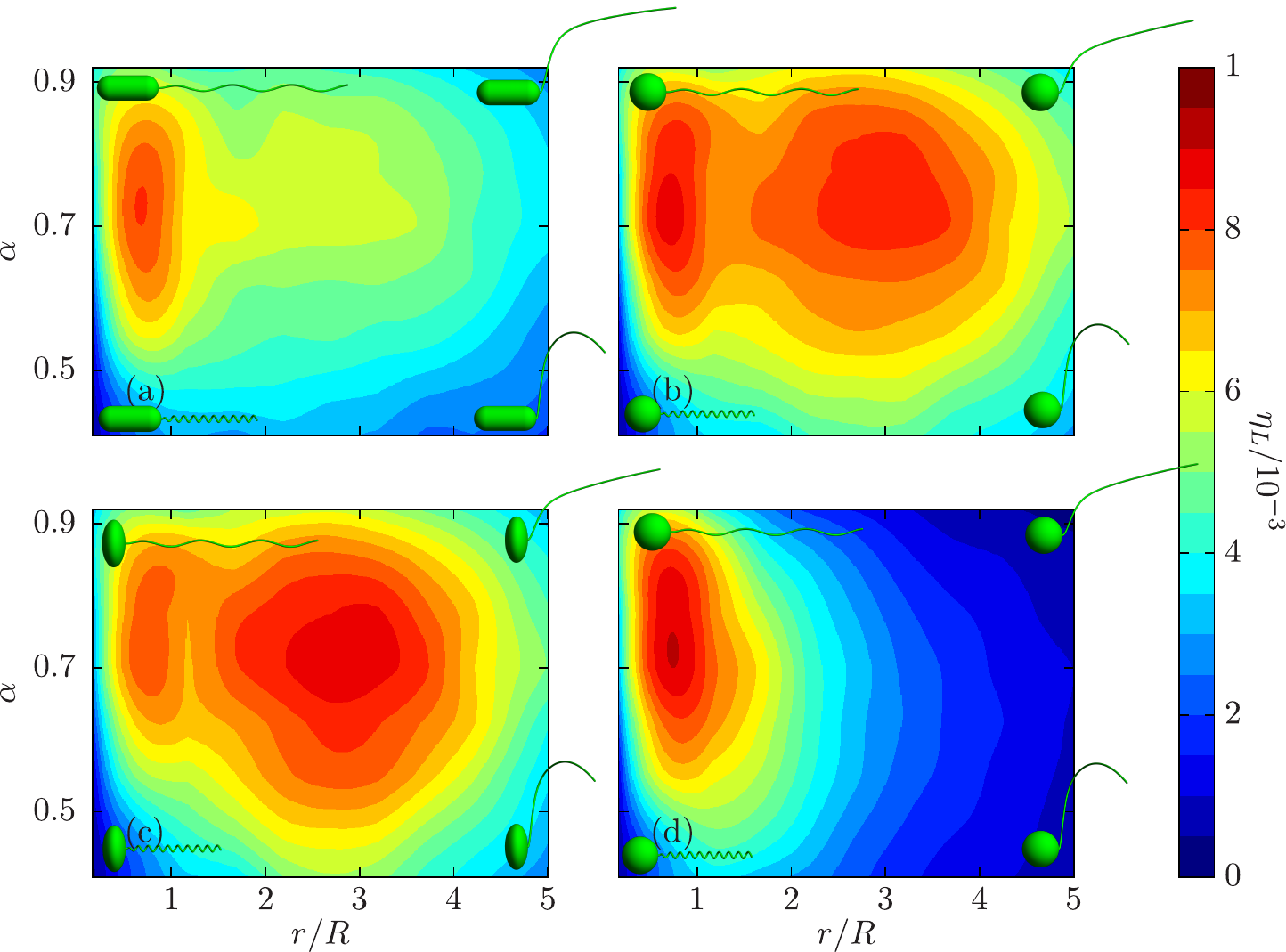} 
   \caption{(color online) Efficiency $\eta_L$ as a function of helical radius $r$ and pitch parameter $\alpha$ (see Fig. \ref{fig:helixswimmers} and text) for body aspect ratio $L/D = 2.5$ (a), $L/D =1.0$ (b), and $L/D = 0.5$ (c). Panel (d) shows the effiency for $L/D=1$ calculated with the simplified $3 \times 3$ resistance tensor, featuring only a single local maximum. The swimmers shown in the four corners of each panel further illustrate the shapes covered in the $r$-$\alpha$ plane.}
   \label{fig:helixswimmerplots}
\end{figure*} 
 
In Fig. \ref{fig:helixswimmerplots} we show the dependence of $\eta_L$ on the flagellum radius $r$ and the pitch parameter $\alpha$, for three different cell body aspect ratios $L/D=2.5$ (a), $1$ (b) and $0.5$ (c). The flagellum length and radius are fixed at $\ell/R = 11$ and $\rho/R = 0.051$, corresponding to the values for E. coli.

For an E. coli-like cell body with $L/D =2.5$, we find in Fig. \ref{fig:helixswimmerplots}(a) a single maximum $\eta_L = 0.0089$ at $r/R = 0.68$ and $\alpha = 0.80$. This shape is shown in Fig. \ref{fig:helixswimmers}(b). 
Surprisingly, for smaller $L/D$ a second maximum develops, as can be seen for a spherical body with $L/D=1$ in Fig. \ref{fig:helixswimmerplots}(b), with a local optimum of $\eta_L =0.0085$ for a `wagging tail'-like shape at $r/R = 3.1$ and $\alpha =0.75$ (Fig. \ref{fig:helixswimmers}(d)), next to the global optimum of $\eta_L = 0.0089$ for $r/R = 0.68, \alpha = 0.75$ (Fig. \ref{fig:helixswimmers}(c)).
In Fig. \ref{fig:helixswimmerplots}(d), we show $\eta_L$ for $L/D=1$, but now calculated with a simplified $3 \times 3$ resistance tensor, where the off-diagonal hydrodynamic interactions between cell body and flagellum are ignored. We observe (from comparison with Fig. \ref{fig:helixswimmerplots}(b)) that although this approximation produces fairly accurate results in the small-$r$ regime, it is unable to reproduce the second local maximum of the `wagging tail'-type swimmer at larger $r$. This is in agreement with the observation from the animations that this shape shows a large transversal motion (or `wobble'), indicating that these transversal degrees of freedom are not negligible.
To calculate the efficiency of even smaller $L/D$, we consider a cell body of an oblate ellipsoid of $L/D =0.5$. In Fig. \ref{fig:helixswimmerplots}(c), we observe that  the `wagging tail' local maximum becomes a global maximum, with $\eta_L = 0.0084$ for $r/R = 2.7$ and $\alpha = 0.68$ as shown in Fig. \ref{fig:helixswimmers}(e). 
Not shown here are results for $L/D > 2.5$, which we find to be qualitatively similar to the $L/D = 2.5$ case. 
\textcolor{black}{
Neither shown here are results obtained by varying $\rho$ and $\ell$ and fixing the radius and pitch at the values for E. coli ($r/R = 0.33$ and $\alpha = 0.87$). Here, we find that the efficiency increases monotonically with decreasing $\rho$, while as a function of $\ell$ it shows a broad maximum around $\ell/R = 11$. 
We point out that by varying only the two shape parameters $r$ and $\alpha$ in Fig. \ref{fig:helixswimmerplots}, we found maxima that are not (global) maxima in the full five-dimensional shape parameter space. A five-dimensional optimization, which may be material for future work, could result in obtaining either a single global maximum or several local maxima.
}

We find the optimal radius and pitch parameter of Fig.\ref{fig:helixswimmerplots}(a) and (b) to be in agreement with the results of earlier optimization studies on similarly (not identically) shaped swimmers using resistive force theory \cite{higdon1979hydrodynamics} or boundary element methods \cite{fujita2001optimum,shum2010modelling}. However, none of these studies report the second optimal `wagging tail'-type flagellum, while it does resemble the optimal (externally driven) swimmer calculated in Ref. \cite{keaveny2013optimization}, which also exhibits only a single maximum. Interestingly, this type of flagellated swimmer is (to our knowledge) not observed in nature. 

\section{Summary \& Outlook}
In summary, in this work we have set up a method that combines a theoretical framework for the equations of motion of an $N$-component swimmer, with numerical bead-shell model calculations. This method allows for the calculation of the displacement and efficiency of any general-shaped swimmer, with relatively short computation time.
First we employed this method to calculate the friction coefficients for the platonic solids and found that the hydrodynamic radius may be estimated by $\sqrt{A/4\pi}$.

When applied to the class of three-body swimmers, we found that for long arms the displacement and efficiency are determined by the single-body friction coefficient, while maximally efficient strokes are performed when the bodies can approach as closely as possible. Next, we have applied this scheme to a swimmer with a helical flagellum, modelled after an E. Coli bacterium. The calculated swimming velocity and body/flagellum rotation rates are in fairly good agreement with the measured values for E. Coli. Also, the swimming efficiency shows an intricate dependency on the swimmer geometry. Within this class of swimmers, we distinguish two types of efficient swimming flagella: a helical flagellum that resembles the flagellum of the E. Coli bacterium and a stretched `wagging tail'-type flagellum, where this second optimal shape is not reported in earlier optimization studies. 

Our theoretical description can straightforwardly be extended to a single swimmer close to a wall by exploiting analogies to image charge effects \cite{kwaadgras2014orientation,swan2007simulation}. 
This is a natural next step given the fact that many experiments are conducted in a quasi two-dimensional geometry. Hydrodynamic pair interactions, and perhaps even many-body interactions, can also be accounted for, albeit at the expense of numerical effort. Work in these directions is being pursued.

\section{Supplementary Material}
See supplementary material for animations of the swimmers considered in this text.

\section{Acknowledgements}
This work is part of the D-ITP consortium, a program of the Netherlands Organisation for Scientific Research (NWO) that is funded by the Dutch Ministry of Education, Culture and Science (OCW). We acknowledge financial support from an NWO-VICI grant.

\providecommand{\noopsort}[1]{}\providecommand{\singleletter}[1]{#1}%

\end{document}